\documentclass{iaus}
\usepackage{graphicx}

\title[Precision Astrometry and Local Dark Matter] 
{Precision Astrometry, Galactic Mergers, Halo Substructure and Local Dark Matter}

\author[Steven R. Majewski]   
{Steven R. Majewski$^1$}

\affiliation{$^1$Department of Astronomy, University of Virginia\\
P.O. Box 400325, Charlottesville, VA, 22904-4325, U.S.A. \\ email: {\tt srm4n@virginia.edu}}

\pubyear{2008}
\volume{248}  
\pagerange{1}
\jname{A Giant Step: from Milli- to Micro-arcsecond Astrometry}
\editors{W. Jin, ed.}
\begin{document}

\maketitle

\begin{abstract}
The concordance Cold Dark Matter model for the formation of structure in the Universe, 
while remarkably successful at describing observations on large scales, has a number of
problems on galaxy scales.  The Milky Way and its satellite system provide a key laboratory for 
exploring dark matter (DM) in this regime, but some of the most definitive tests of local DM
await microarcsecond astrometry, such as will be delivered by the Space Interferometry Mission
(SIM Planetquest).
I discuss several tests of Galactic DM enabled by future microarcsecond astrometry.

\keywords{astrometry, stars: kinematics, 
Galaxy: halo, 
Galaxy: kinematics and dynamics, 
galaxies: interactions, 
dark matter}
\end{abstract}

\firstsection 
\section{Introduction}
Since the seminal study of Searle \& Zinn (1978) the notion of accretion of ``subgalactic units", 
including ``late infall", has been a central concept of stellar populations studies.  
N-body simulations of the formation of structure in the Universe in the
presence of dark matter (and dark energy) also show galaxies (and all large
structures) building up hierarchically.  
However, while the active merging history on
all size scales demonstrated by high resolution, Cold Dark Matter (CDM)
numerical simulations have had remarkable success in matching the
observed properties of the largest structures in the Universe (like
galaxy clusters), they are a challenge to reconcile with the observed
properties of structures on galactic scales.  The Milky Way (MW) and its
satellite system are a particularly important laboratory for testing
specific predictions of the CDM models.  The era of microarcsecond
astrometry such as will be delivered by SIM Planetquest, 
will enable a number of definitive dynamical tests of local 
CDM by way of determining the distribution of Galactic dark matter (DM).
We discuss several of these tests here.

\section{Measuring $\Theta_{\rm LSR}$ with the Sagittarius Stream}

An understanding of the distribution and amount of DM in the Galaxy --- indeed,
virtually every problem in Galactic dynamics --- depends on establishing
the benchmark parameters $R_0$, the solar Galactocentric
distance, and $\Theta_{\rm LSR}$, the Local Standard of Rest (LSR) velocity.  A 
3\% error in both $R_0$ and $\Theta_{\rm LSR}$ leads to a 5\% error in the 
determination of the Galactic mass scale using, e.g., traditional Jean's equation methods.

Despite decades of effort, the rate of Galactic rotation at the Sun's position 
remains uncertain, with measurements varying by 20\% or more.  
Hipparcos proper motions have been used to determine that $\Theta_{\rm LSR}
=(217.5\pm 7.0) (R_0/8)$ km s$^{-1}$ using disk Cepheid variables 
(Feast \& Whitelock 1997) and $(240.5\pm 8.3) (R_0/8)$ km s$^{-1}$
from OB stars (Uemura et al. 2000), while Hipparcos proper motions
of open clusters have yielded $207 (R_0/8)$ km s$^{-1}$
(Dias \& L{\'e}pine 2005) [though see the alternative analysis 
of these open cluster by Frinchaboy 2006 and Frinchaboy \& Majewski 2006,
which yields $(221\pm3)$ km s$^{-1}$ for any $7 \le R_0 \le 9$]. 
Meanwhile, Hubble Space Telescope measurements of the proper motions
of bulge stars against
background galaxies in the field of the globular cluster M4
yield $\Theta_{\rm LSR}=(202.4\pm20.8) (R_0/8)$ km s$^{-1}$ (Kalirai et al. 2004) and
$(220.8 \pm 13.6) (R_0/8)$ km s$^{-1}$ (Bedin et al. 2003).
Radio measures of the proper motion of Sgr A$^*$
(Reid \& Brunthaler 2004) yield 
a transverse motion of $(235.6\pm1.2) (R_0/8)$ km s$^{-1}$,
but, when corrected for the solar peculiar motion, yield 
$\Theta_{\rm LSR}=220 (R_0/8)$ km s$^{-1}$ (M. Reid, this proceedings).
%
Of course, as may be seen, most of these measures 
depend on an accurate measure of $R_0$ as well as knowledge of the solar motion
(which is apparently now more controversial than previously suspected; 
cf. the significantly smaller solar motion derived by Dehnen \& Binney 1998
compared to that from M. Reid, this proceedings).  
Moreover, considerations of non-axisymmetry of the disk yield corrections to
the measurements that suggest
 $\Theta_{\rm LSR}$ may be as low as $184 \pm 8$ km s$^{-1}$ (Olling \& Merrifield 1998)
or lower (Kuijken \& Tremaine 1994).

Based on recent estimates of the luminosity of the MW (Flynn et al. 2006), any
$\Theta_{\rm LSR}$ exceeding 220 km s$^{-1}$ puts our Galaxy more than 1$\sigma$
away from the Tully-Fisher relation.  
Clearly, additional independent methods to ascertain $\Theta_{\rm LSR}$ are valuable for establishing 
with certainty the MW mass scale and 
whether our home galaxy is unusual.  

An ideal method for ascertaining $\Theta_{\rm LSR}$ would be to measure
the solar motion with respect to a {\it nearby} reference known to be
at rest in the Galactic reference frame (at least in the disk rotation direction), 
since this would not only 
overcome traditional difficulties with working in the highly dust-obscured and crowded
Galactic center, but be
independent of $R_0$ and any assumptions
that the reference lies in the center of the MW potential.  The debris stream
from the tidally disrupting Sagittarius (Sgr) dSph galaxy
provides almost nearly this ideal situation.  This dSph and its
extended tidal debris system orbit the MW in almost a polar orbit, with a Galactic 
plane line-of-nodes
almost exactly along the Galactic $X$-axis.  The remarkable
coincidence that the Sun presently lies within a kiloparsec
of the Sgr debris plane, which has 
a pole at $(l_p,b_p)=(272, -12)^{\circ}$ (Majewski et al. 2003), means
that the motions of Sgr stars {\it within} this plane are almost entirely contained in their
Galactic $U$ and $W$ velocity components, whereas their $V$ motions almost entirely
reflect {\it solar motion}.  Thus, in principle, the solar motion can be derived directly
by the $\mu_l \cos(b)$ dimension of the Sgr stream proper motion  --- though
refined results come from comparing to Sgr models that can account for deviations from the ideal case
(Majewski et al. 2006).  
An additional advantage is that stars
in the Sgr stream, particularly its substantial M giant population (Majewski et al. 2003),
are ideally placed for uncrowded field astrometry at high
MW latitudes, and at relatively bright magnitudes.  

This combination of factors means this experiment requires only the most
modest precisions from SIM, is well within the capabilities of Gaia, and may even be within
the grasp of future high quality, ground-based astrometric studies.  
The latter was demonstrated by ground-based astrometry obtained by
Casetti-Dinescu et al. (2006) in two Kapteyn Selected Area fields positioned
on the Sgr trailing arm, where dozens of 1-3 mas year$^{-1}$ proper motions for
$V=19-21$ main sequence stars in the Sgr stream were averaged to derive a mean
motion; these data showed that with more stars and more fields at this level
of astrometric precision one might hope to distinguish MW models at the level of 10s of km s$^{-1}$.
To achieve $\sim1$ km s$^{-1}$ precision in the solar motion with this method will require
$\sim0.01$ mas year$^{-1}$ {\it mean} proper motions for sections of the Sgr trailing arm ---
a result that will be trivial for microarcsecond level, space astrometry.

\section{Probing the 
Galactic Potential with Tidal Streams}

It is now well known that the Galactic halo is inhomogeneous and coursed
by streams of debris tidally pulled out of accreted satellites.
Standard methods of measuring the
MW's gravitational potential using a tracer
population whose orbits are assumed to be random and well-mixed are
systematically biased under these circumstances
(Yencho et al. 2006). 
However, these streams themselves provide uniquely sensitive probes of the
MW potential.  If we could measure the distances, angular positions,
radial velocities and proper motions of debris stars, we could
integrate their orbits backwards in some assumed Galactic potential.
Only in the correct potential will the path of the stream stars ever
coincide in time, position and velocity with that of the parent satellite
(see Fig.\ 1).

\begin{figure}[b]
\begin{center}
\begin{center}
\includegraphics[width=5.0in]{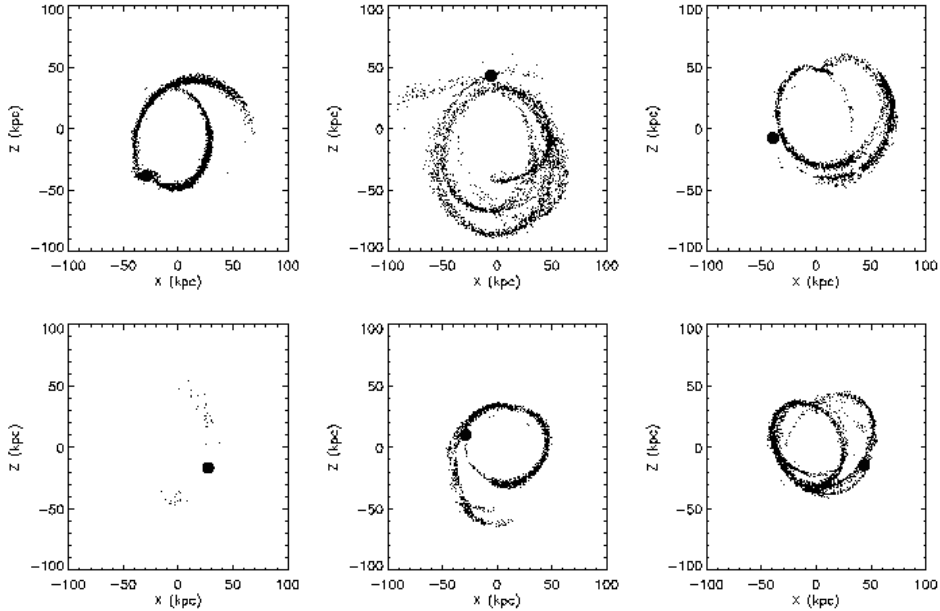} 
\end{center}
\caption{A demonstration of the use of tidal streams to measure the shape
and strength of the Galactic potential
if precise 6-D information is available for stream stars. 
A Sgr-like tidal stream was
  created by the disruption of a dwarf satellite in a rigid Galactic
  potential through a semi-analytical N-body simulation (e.g., as
  in Law et al. 2005; panel a).
With complete 6-D phase space information
  in hand, guesses may be
  made on the strength and shape of the Galactic potential, and the
  orbits of the individual stars in the tidal streams run backwards
  under these assumed potentials.  Panels (b) and (c) demonstrate what
  happens when the strength of the Galactic potential is underestimated
  by varying degrees: When the orbits are run backwards, the tidal
  stream stars orbit at too large a radius and do not converge to a
  common phase space position.  In panels (e) and (f) the strength of
  the Galactic potential has been overestimated, and when the clock is
  run backwards the tidal stream stars assume orbits that are too small
  and once again do not converge on a common phase space position.  In
  panel (d) a Galactic potential of the correct strength was guessed,
  and, when the stream star orbits are run backwards, the tidal stream
  stars collect back into the core of the parent satellite (shown as the
  dot in all panels).  }
\label{backwards_stream}
\end{center}
\end{figure}

Microarcsecond proper motions and parallaxes combined with ground-based
radial velocities provide
everything needed to undertake the experiment.  
If we find a coherent stream without an associated satellite, the same
techniques apply, but with the parent satellite's position and velocity
as additional free parameters.  The usefulness of Galactic tidal streams
for probing the MW potential has long been recognized, and the
promise of astrometric space missions to provide the last two dimensions of
phase space information for stream stars 
eagerly anticipated
(e.g., Johnston et al. 1999, Pe{\~n}arrubia et al. 2006).

Applying this method to simulated data observed with the $\mu$as
yr$^{-1}$ precision proper-motions possible with SIM and km s$^{-1}$
radial velocities suggests that 1\% accuracies on Galactic parameters
(such as the flattening of the potential and circular speed at the Solar
Circle) can be achieved with tidal tail samples as small as 100 stars
(Johnston et al. 1999, Majewski et al. 2006).  Dynamical friction is not an important
additional consideration if the change in the energy of the satellite's
orbit in $N_{\rm orb}$ orbits is less than the range in the energies of
debris particles.  For $N_{\rm orb}=3$, this condition is met for all
satellites except the Magellanic Clouds and Sgr.  Evolution of the
Galactic potential does not affect the current positions of tidal
debris, which respond adiabatically to changes in the potential and
therefore yield direct information on the {\it present} Galactic mass
distribution independent of how it grew (Pe{\~n}arrubia et al. 2006).

It is important to sample tidal tails
to points where stars were torn from the satellite at least one radial
orbit ago and have thereby experienced the full range of Galactic
potential along the orbit (Johnston 2001).  The
stars also need to have proper motions measured sufficiently accurately
that the difference between their own and their parent satellite's
orbits are detectable --- this translates to requiring proper motions of
order 100 $\mu$as yr$^{-1}$ for the Sgr stream but 10s of $\mu$as yr$^{-1}$ for
satellites farther away.
Ideally, we would like to probe stars from different tidal tails at a variety of 
Galactocentric distances and orientations with respect to the Galactic
disk to derive the shape and density profile of the DM. 
In the last few years evidence for dSph galaxy tidal tails has
been discovered around systems at large Galactic radii 
(Mu\~noz et al. 2006a, Sohn et al. 2007)
and, with SIM, extended versions of such distant streams can be used to
trace the Galactic mass distribution as far out as
the virial radius with an unprecedented level of detail and accuracy.

A key problem that is currently debated but that will easily be resolved
with precision astrometry of tidal tails is the shape of the Galactic halo.
CDM predicts that host DM halos should be oblate (minor to
major axis ratio, $q\sim0.7$) with flattening increasing with radius,
whereas warm and self-interacting DM models tend to find rounder halos
(Bullock et al. 2002).  Studies of the halos of external galaxies --- whose
shapes can be measured via flaring of HI disks, the shapes of diffuse
x-ray isophotes (for ellipticals), or with polar rings --- typically
find rather oblate results (see summaries in Combes 2002,
Merrifield 2005).  
Meanwhile a controversy has
emerged regarding the true shape of the halo as measured by the Milky
Way's own polar ring system --- the tidal stream of the Sgr
dwarf spheroidal galaxy.  Depending on how the extant data are analyzed
and modeled the debris of the Sgr dwarf has been argued to be
consistent with a rather prolate ($q\sim1.4$; Helmi 2004), near
spherical (Ibata et al. 2001, Fellhauer et al. 2006), slightly oblate
($q\sim0.93$; Johnston et al. 2005, Law et al. 2005) or oblate
($q=0.85$; Martinez-Delgado et al. 2004) Galactic halo.  The definitive
measurement of the shape of the Galactic halo within the Sgr
orbit will come from SIM proper motions of stars in its tails, 
and the variation of the halo shape with radius will
come by repeating the technique with other tidal streams at other
distances.  Interestingly, recent measures of the mass profile of the
MW using stellar tracers suggest that it gets rounder with radius
(Kuijken 2003), in contradiction to the predictions of CDM.  

The possible existence of a significant fraction
of the halo in the form of dark satellites has been
discussed in recent years
(Moore et al. 1999, Klypin et al. 1999).
These putative dark subhalos could
scatter stars in tidal tails, possibly compromising their use as
large-scale potential probes, but astrometric measurements of stars in
these tails could, on the other hand, be used to assess the importance
of dark substructure (Ibata et al. 2002, Johnston et al. 2002).  Early tests of such
scattering using only radial velocities of the Sgr stream suggest a
MW halo smoother than predicted (Majewski et al. 2004), but
this represents debris from a satellite with an already sizable
intrinsic velocity dispersion.
Because scattering from subhalos should be most
obvious on the narrowest, coldest tails (such as those seen from globular clusters, like
Palomar 5 -- e.g., Grillmair \& Dionatos 2006)
these could be used to probe the DM substructures, whereas the
stars in tails of satellite galaxies such as Sgr, with larger
dispersions initially and so less obviously affected, can still be used
as global probes of the Galactic potential.

Gaia will allow these tests to be attempted for nearby streams, whereas with SIM, 
it will be possible for the first time to probe the full
three-dimensional shape, density profile, extent of and substructure
within our closest large DM halo.

\section{Probing the Galactic Potential with Hypervelocity Stars}

A complementary method for sensing the shape of the Galactic potential can
come from SIM observations of hypervelocity stars (Gnedin et al. 2005).
Hills (1988) postulated that such stars would be ejected at speeds exceeding 1000 km s$^{-1}$
after the disruption of a close binary star system deep in the
potential well of a massive black hole, but HVSs can also be produced
by the interaction of a single star with a binary black hole
(Yu \& Tremaine 2003).  
Recently Brown et al. (2006) report on five stars with Galactocentric velocities
of 550 to 720 km s$^{-1}$, and argued persuasively that these must be
``unbound stars with an extreme velocity
that can be explained only by dynamical ejection associated with a
massive black hole". After the success of these initial surveys, 
it is likely that many more HVSs will be discovered
in the next few years.

If these stars indeed come from the Galactic center, the orbits are tightly
constrained by knowing their point of origin.  In this case, as Gnedin et al. (2005)
demonstrate, the non-spherical shape of the Galactic
potential --- due in part to the flattened disk and in part to the
triaxial dark halo --- will induce non-radial inflections (which will be primarily in transverse 
direction at large radii) in the velocities of the HVSs of order
5--10 km s$^{-1}$, which corresponds to 10--100 $\mu$as yr$^{-1}$. 
Each HVS thus
provides an
independent constraint on the potential, as well as on the solar
circular speed and distance from the Galactic center.
The magnitudes of the known HVSs range from 16 to
20, so their proper motions will be measurable by SIM with an accuracy
of a few $\mu$as yr$^{-1}$, which should define the 
orientation of their velocity vectors to better than 1\%.
With a precision of 20 $\mu$as yr$^{-1}$ the orientation of the 
triaxial halo could be well-constrained and at 10 $\mu$as yr$^{-1}$
the axial ratios will be well-constrained (Gnedin et al. 2005).

\section{Dark Matter within Dwarf Galaxies}

Dwarf galaxies, and particularly dSph galaxies,
are the most DM-dominated systems known to exist.
Relatively nearby Galactic dSph satellites provide the opportunity to study the
structure of DM halos on the smallest scales, and, with microarcsecond astrometry, 
make possible
a new approach to determining the physical nature of DM (Strigari et al. 2007, 2008).
CDM particles have negligible
velocity dispersion and very large central phase-space density, which
results in cuspy density profiles over observable scales
(Navarro et al. 1997, Moore et al. 1998), whereas, 
in contrast, Warm Dark Matter (WDM) has smaller central
phase-space density so that constant central cores develop in the density profiles. 
Because of their small size,
if dSph cores are a result of DM physics then the cores
occupy a large fraction of the virial radii, which makes these particular cores 
more observationally accessible than those in any other galaxy type.
Using dSph central velocity dispersions, earlier constraints on
dSph cores have excluded extremely warm DM,
such as standard massive neutrinos (Lin \& Faber 1983, Gerhard \& Spergel 1992).
More recent studies of the Fornax dSph provide strong constraints
on the properties of sterile neutrino DM
(Goerdt et al. 2006, Strigari et al. 2006).

The past decade has seen substantial progress in measuring radial
velocities for large numbers of stars in nearby dSph galaxies, and
the projected radial velocity dispersion profiles are generally found to be 
roughly flat as far out as they can be followed (Mu\~noz et al. 2005, 2006a, 
Walker et al. 2006a,b, 2007, Sohn et al. 2007, Koch et al. 2007a,b, Mateo et al. 2007).
By combining such radial velocity profiles with the surface density
distributions of dSph stars (which are well-fitted by King profiles, modulo
slight variations, especially at large radii),  
it is typical to derive DM density profiles
by assuming dynamical equilibrium and solving the
Jeans equation (e.g, Richstone \& Tremaine 1986). 
The results of such equilibrium analyses typically imply
at least an order of magnitude more dSph mass in dark
matter than in stars as well as 
mass-luminosity ratios that increase with radius
 --- in some cases quite substantially (e.g., Kleyna et al. 2002) ---
though at large radii tidal effects probably complicate this picture
(Kuhn 1993, Kroupa 1997, Mu\~noz et al. 2006a, 2007, Sohn et al. 2007, Mateo et al. 2007),
and in some cases mass-follows-light models incorporating tidal disruption describe the observations
quite well (Sohn et al. 2007, Mu\~noz et al. 2007).
Knowing whether mass follows light in dSphs or if the luminous components
lie within large extended halos is critical to establishing the
regulatory mechanisms that inhibit the formation of galaxies in
all subhalos.

Unfortunately, as shown by Strigari et al. (2007), equilibrium model solutions to
the Jeans equation are degenerate and satisfied by DM 
density profiles with either cores or cusps.
This is because there is a strong degeneracy between the inner slope
of the DM density profile and the velocity anisotropy,
$\beta$, of the stellar orbits, which leads to a strong dependency
of the derived dSph masses on $\beta$.
Radial velocities alone cannot break this
degeneracy (Fig.\ 2)
even if the radial velocity samples are
increased to several thousand stars (Strigari et al. 2007).
The problem is further compounded if we add triaxiality, Galactic
substructure, and dSph orbital shapes to the allowable range of
parameters.  The only way to break the mass-anisotropy degeneracy
is to measure more phase space coordinates per star,
and in particular to acquire transverse velocities for
the stars, because the Jeans equation solved in the transverse
dimension probes the anisotropy differently
than in the radial velocity dimension.
Thus, combining high precision proper motions with even the present
samples of radial velocities
holds the prospect to break the anisotropy-inner slope degeneracy (Fig.\ 2).

\begin{figure}[ht]
\begin{center}
\hfill\includegraphics[width=0.4\textwidth]{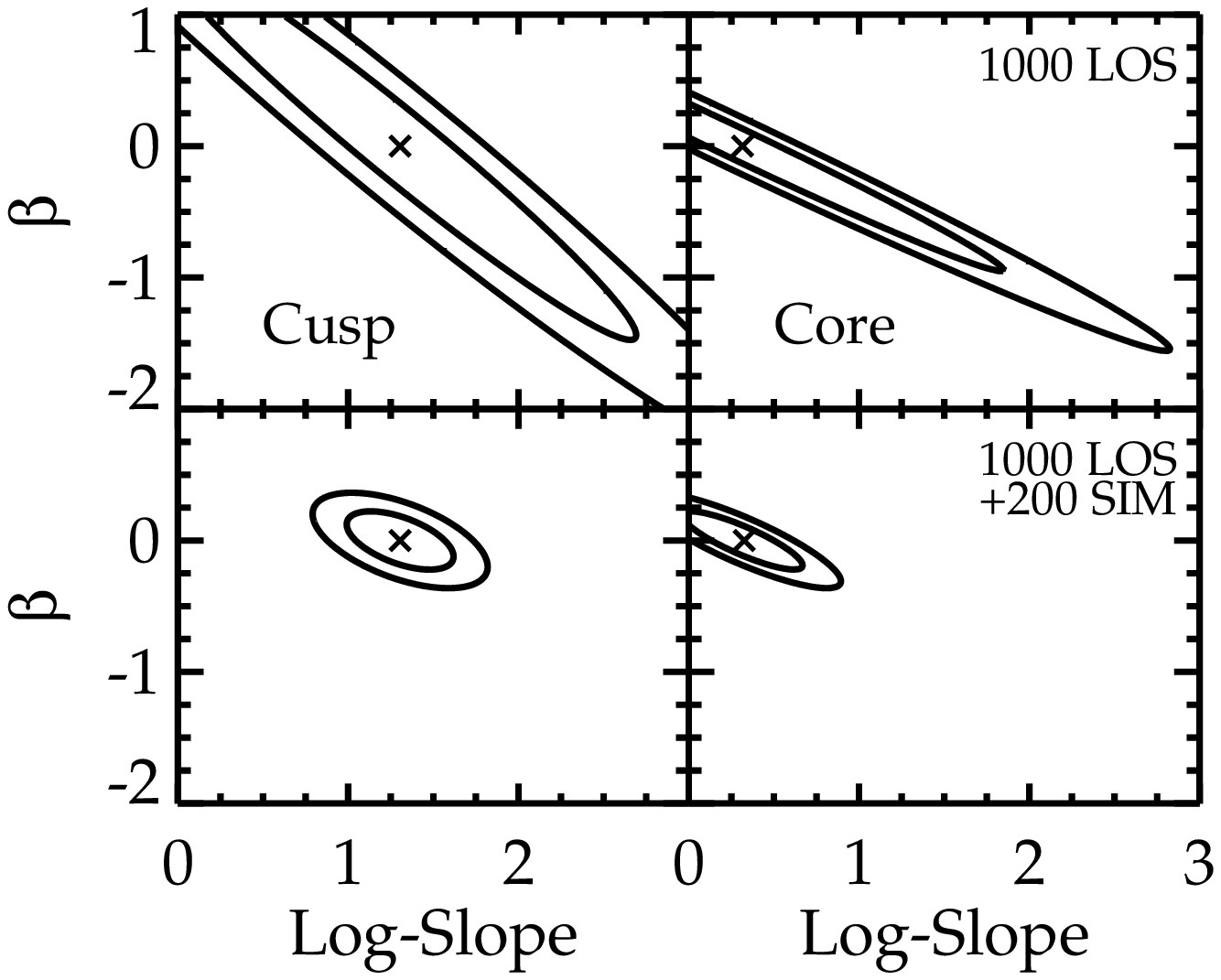}\hfill  
\includegraphics[width=0.4\textwidth]{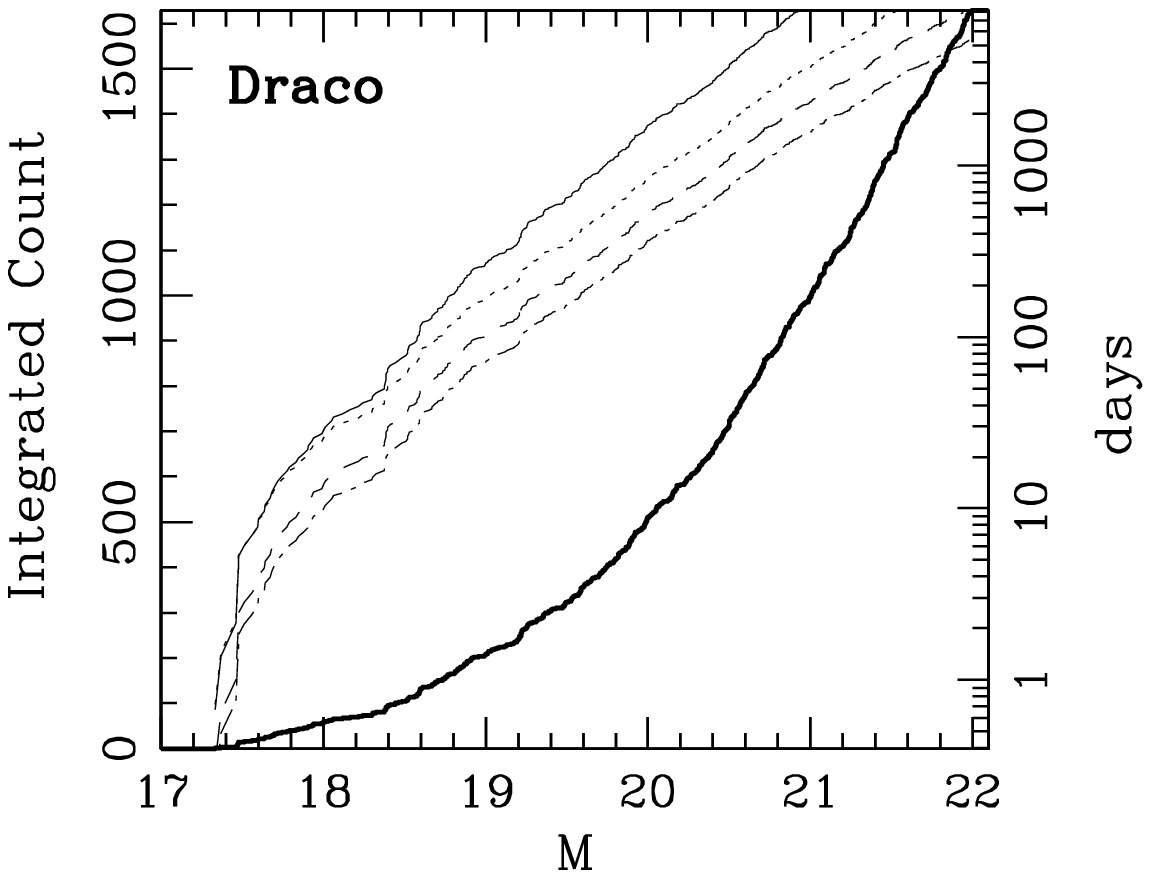}\hfill\\
\end{center}
\caption{{\bf Left panels:} A demonstration of the ability to recover information
on the nature of DM using observations of dSph stars,
from analytical modeling by Strigari et al. (2007).  Ellipses indicate
the $68\%$ and $95\%$ confidence regions for the errors in the
measured dark halo
density profile slope (measured at twice the King core radius) and
velocity anisotropy parameter $\beta$ in the case where only
radial velocities are available for 1000 stars in a particular
dSph (top panels).  A significant improvement is derived from 
the addition of 200 SIM proper motions providing
5 km s$^{-1}$ precision transverse velocities (bottom panels).
The left (right) panels correspond to a cusp (core) halo model for
dSphs and the small x's indicate the fiducial, input model values.
}
\label{BETA_SLOPE}
\caption{{\bf Right panel:} Potential SIM PlanetQuest exploration of the Draco dSph 
would need to probe to $V \sim M=19$ to derive a sample of 200 red giants
as seen by its Washington $M$-band luminosity function ({\it thick line and left axis}).
The thin lines represent the number of days ({\it right axis}) necessary to observe
all the stars to a given magnitude limit with SIM and for a
given transverse 
velocity uncertainty: 3 km s$^{-1}$ ({\it thin solid line}),  5 km s$^{-1}$ ({\it dotted line}), 7 km s$^{-1}$ 
({\it dashed line}) and 10 km s$^{-1}$ ({\it dotted-dashed line}).  From Strigari et al. (2008).
}
\label{Majewski_fig3}
\end{figure}

The most promising dSphs to obtain proper motions for will be
the nearby (60-90 kpc distant) 
systems Sculptor, Draco, Ursa Minor, Sextans and Bootes, which 
include the most DM-dominated
systems known (Mateo 1998, Mu\~noz et al. 2006b, Martin et al. 2007) 
as well as a system with a more modest $M/L$ (Sculptor).
To sample the velocity dispersions properly will require
proper motions of $> 100$ stars per galaxy with accuracies of
7 km s$^{-1}$ or better (less than 15 $\mu$as yr$^{-1}$).
Strigari et al. (2007) show that with about 200 radial velocities and 200 transverse velocities
of this precision
it will be possible to reduce the error on the log-slope of the dark
matter density profile to about 0.1 --- which is
an order of magnitude smaller than the errors attainable from a sample
of 1000 radial velocities alone, and sensitive enough to rule out nearly
all WDM models (Fig.\ 2).
Figure 3 shows that exploring even these nearby dSph
systems requires precision proper motions of stars to $V \gtrsim 19$, 
a task well beyond the capabilities of GAIA, but well-matched to the
projected performance of SIM, though requiring a Key Project level
of observing time (e.g., 100 days of SIM observing for 200 Draco stars).
Note that while the Sgr dSph is several times closer than these
other dSphs, it is obviously a system in dynamical non-equilibrium;
exploring this system will be of interest as a case study for 
establishing the internal dynamical effects
of tidal interaction on dSphs.

 
\acknowledgements
I appreciate assistance from James Bullock, Jeffrey Carlin, Peter Frinchaboy,
Kathryn Johnston, Ricardo Mu\~noz, Richard Patterson, Louie Strigari, 
and Scott Tremaine in preparation of this presentation and contribution.

\end{document}